# Carrier mediated reduction of stiffness in nanoindented crystalline Si(100)

S. Kataria,[†] Sandip Dhara,[*] S. Dash, A. K. Tyagi

*Surface and Nanoscience Division, Indira Gandhi Centre for Atomic Research. Kalpakkam-603102, India*

*Abstract*

We report the observation of carrier mediated decrease in the stiffness of crystalline (*c*)-Si(100) under nanoindentation. The apparent elastic modulii of heavily doped (~$1 \times 10^{21}$ cm$^{-3}$) *p*- and *n*-type *c*-Si are observed to be lower by 5.3—7.5% than the estimated value for intrinsic (~$1 \times 10^{14}$ cm$^{-3}$) *c*-Si. The deviation observed with respect to elastic modulus remarkably matches with the estimated value while considering the electronic elastic strain effect on carrier concentration as an influence of negative pressure coefficient of band gap for Si ($\Gamma$-$X$). The value is predominantly higher than the reported value of a decrease of 1-3% in stiffness as an effect of impurity in *c*-Si.

[*]Corresponding author Email : dhara@igcar.gov.in

[†] Presently at University of Siegen, Graphene-Based Nanotechnology, Hölderlinstr. 3, Siegen, Germany, Email : s_kataria2k2@yahoo.co.in

## 1. Introduction

It is well known that the strain is a normalized measure of deformation representing the displacement between particles in the body relative to a reference length. Photoexcited elastic deformation, thus can be influenced by thermoelastic, and electronic strain (electronic deformation) in semiconducting materials. The former component is the consequence of the sample surface displacement due to the localized photon assisted heating and the later one is the consequence of the sample surface displacement when an electron-hole pair is generated.[1] The electronic strain was reported to play an important role in the mechanical deformation of semiconducting nanostructures under illumination with photon energy greater than the band gap of the semiconductor, demonstrating the significance of mechanical, optical, and electronic coupling.[2] Incidentally, the thermoelastic effect will be also prominent under photoexcitation above band gap where all the energies will be absorbed. The effect will be particularly strong in case of indirect band gap semiconductors where electron-hole recombination process occurs with non-radiative transitions involving energy transfer to phonon leading to the increase in lattice temperature and affecting the thermoelastic property of the material. At the same time, the elastic deformation due to electronic strain can also be demonstrated in the absence of photoexcitation for indirect band gap semiconductors with different carrier concentration at different doping level in the absence of pre-dominant effect of impurity in altering the mechanical property.

The optical properties of a semiconducting system, namely, ZnO was reported to be modified by elastic stress or strain.[3-5] However, the inverse effect where the influence of light on the elastic properties of wurtzite (*w*-)ZnO nanowire was reported in nanostructures considering both the surface effect and the electronic strain induced by the photogeneration of free carriers.[6] The study also predicted possible realization of nanoscale optical tunable surface acoustic wave device where the elastic properties of *w*-ZnO nanowire were reported to be tuned by the optical illumination. Diameter dependent mechanical properties of ZnO nanowire is also a matter of recent discussion in this regard.[7] In order to understand the effect of carrier concentration on mechanical properties, we have considered another semiconducting system with varying carrier concentration ranging from intrinsic to heavily doped Si. Mechanical properties of an indirect band gap material also will be relevant in the absence of photoexcitation where thermoelastic effect will be absent, as discussed earlier. The issue is important after publication of the interesting report about photoinduced stiffening of *w*-ZnO,[6] where the authors have emphasized in their report that one can attempt to understand the phenomenon qualitatively from the positive pressure coefficient of band gap,[8] as electronic strain is a function of pressure coefficient for the band gap along with amount of carrier concentration of a system.[2] However, quantitatively it was difficult to understand such a large change in apparent elastic modulus under illumination with excitation above band gap.

Thus, we assess the model in a well established indirect band gap semiconductor, namely, *c*-Si, with varying carrier concentration for both *p*- and *n*-doping cases. Mechanical properties are measured using a nanoindentor. In the absence of thermoelastic component with no photoexcitation on indirect band semiconductor, the effect of excess carrier concentration on electronic strain is parameterized for the calculation of softening of apparent elastic modulus.

## 2. Experimental

Commercial *c*-Si(100) with doping concentration of $1 \times 10^{14}$ cm$^{-3}$ (intrinsic), medium and heavily As-doped *n*-type (~$1 \times 10^{17}$ cm$^{-3}$) and $n^{++}$-type (~$1 \times 10^{21}$ cm$^{-3}$) wafer were used in the present study. We had also used medium and heavily B doped *p*-type (~$2 \times 10^{17}$ cm$^{-3}$) and $p^{++}$-type (~$1 \times 10^{21}$ cm$^{-3}$) wafers for the present investigation.

Nanoindentor (CSM Instruments, Switzerland) fitted with a calibrated Berkovich diamond indenter (inscribed angle 65.3°) tip diameter ≈100 nm was used for the evaluation of mechanical properties. The Berkovich tip is advantageous over conventional tips for its sharp and well-defined tip geometry (spherical surface not always perfect in case of spherical tip) and its role in measuring modulus and hardness values for bulk and thin film sample away from the plastic limit. A Vickers indenter is a four sided pyramidal tip with projected area to depth ratio is also similar to that of a Berkovich tip. However, these indenters are good for very large load work. The loading and unloading rates were kept constant at 50 $\mu N\ s^{-1}$ with a maximum load of 50 mN and holding time of 5 s. Data analysis was performed using Oliver and Pharr method.[9,10] A micro-Raman scattering study was performed with 100X objective with numerical aperture of 0.8 for the analysis of phase transition in *c*-Si under nanoindentation using 514.5 nm excitation of continuous wave $Ar^+$ laser. A monochromator for dispersion was used with 1800 gr/mm grating and the experiment was performed in the backscattering configuration of a spectrometer (inVia Renishaw). A thermoelectric cooled 'back-thinned' CCD detector was used for the detection of scattered intensity.

3. **Results and discussion**

Comparison of nanoindentation on *c*-Si for varying carrier concentration for both *n*- and *p*- types is revealed from the difference in the load-displacement curves (Fig. 1). 'Pop-out' in unloading curve is owing to the phase transition in Si, which is well reported earlier,[11-13] and also evidenced in the present study.

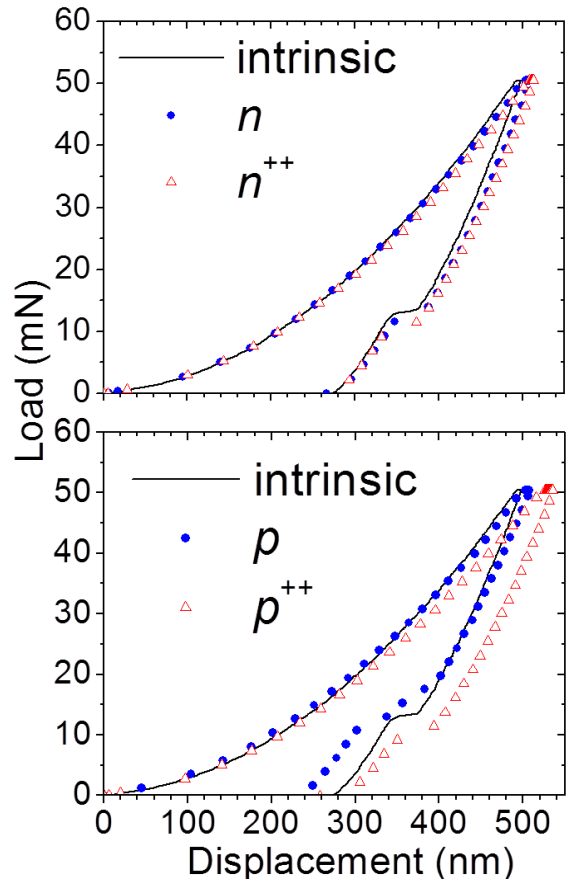

**Fig. 1.** Comparison of load-displacement curves of intrinsic, *n*- and *p*-type *c*-Si by nanoindentations for different carrier concentrations at a load of 50 mN. The observed 'pop-out', during unloading part, is due to the phase transitions occurred in Si during the indentation process as discussed in the text.

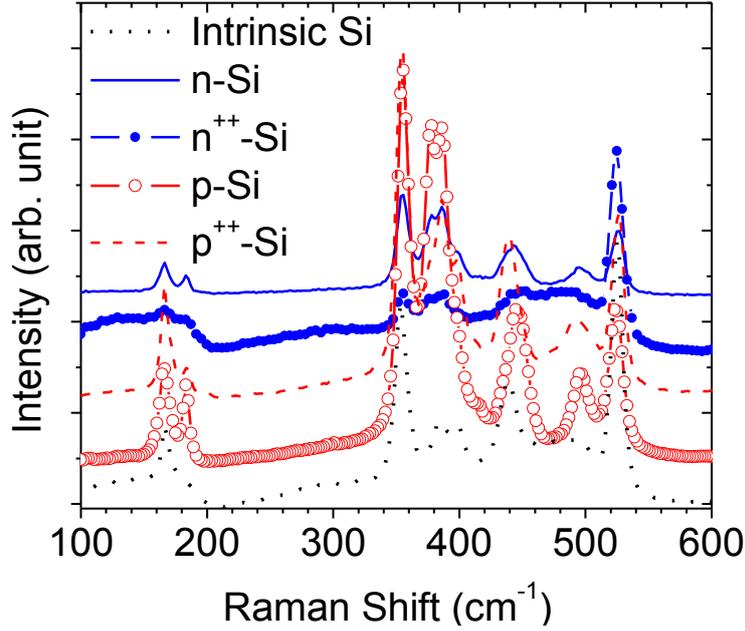

**Fig. 2.** Raman spectra for intrinsic, *n*- and *p*-type *c*-Si of different doping concentration. New Raman peaks are observed in the nanoindented area. These peaks arise from multiple phases of Si (refs. 9-11) which include Si-I (around 525 cm$^{-1}$), Si-III (166, 377, 386, and 438 cm$^{-1}$), Si-XII (353, and 398 cm$^{-1}$) and amorphous (a-) Si (495 cm$^{-1}$). The spectra are shifted vertically for clarity.

Raman peaks around 525 cm$^{-1}$ correspond to Si-I (stress free value at 520 cm$^{-1}$) with expected compressive stress induced shift for nanoindented samples in the high energy side of the spectra (Fig. 2). Peaks around 166, 377, 386, and 438 cm$^{-1}$ correspond to Si-III (10–0 GPa) phase.[12,13] High pressure Si-XII (12–2 GPa) [12,13] phase is also realized for Raman modes at 353, and 398 cm$^{-1}$ along with amorphous Si peak at 495 cm$^{-1}$.[13] We observe that the unloading parts of load-displacement curves (which signify the recovery of elastic deformation as the load is withdrawn) are different in the case of doped Si samples as compared to intrinsic one in both the cases (Fig. 1). As the loads in these samples are similar, diverse elastic deformation means the change in elastic property, which is defined as apparent elastic modulus hereafter. Figure 3 provides a comparison of doping concentration induced softening for both *n*- and *p*-type *c*-Si. The results are derived from the unloading part of the load-displacement curve using the Oliver-Pharr Method.[10,11] We would like to mention here that dopants can alter the mechanical strength especially elastic modulus of Si depending on the atomic size and bond strength between dopant and host Si atoms. However, the additional electronic effects due to induced electronic strain cannot be ignored and are discussed below in the light of doping concentration.

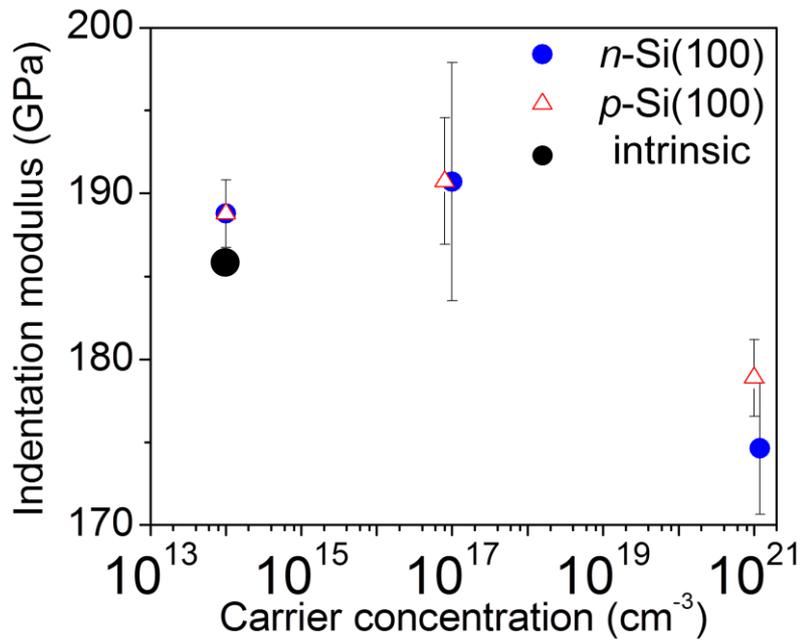

**Fig. 3**. Comparison of the apparent elastic modulus of *n*- and *p*-type *c*-Si by nanoindentations for different carrier concentrations showing a decrease in the stiffness with increasing carrier concentration.

The physical mechanism for the observed softening phenomena can be related to the electronic strain induced by the free charge carriers in heavily doped c-Si. It is well known that the injection of free carriers results in the local mechanical straining of a semiconductor.[6] The electronic strain in the illuminated diamond-type semiconductor is in the form of,[2]

$$S_e = (dE_g/dP)\Delta n/3 \quad \text{------------------------------------------------------------(1)}$$

where $dE_g/dP$ is the pressure coefficient of the band gap energy; $\Delta n$ is the density of the excess charge carriers in cm$^{-3}$. According to Eq. (1), the electronic strain in Si will be negative since the pressure dependence of the band gap energy is negative (-16 meV/GPa ≈ -2.56x10$^{-23}$ cm$^3$).[14] These results in the contraction of the $c$-Si with carrier concentration. The surface contracts and the corresponding deformation is brought about by electronic strain is $S_e$. Hence, the electronic strain as a result of excess free charge carriers adds to the mechanical strain caused by the indent. The net surface response $S_t$ in the heavily doped Si is given by Eq. (2),

$$S_t = S_{\text{intrinsic}} - S_e = (\sigma/E_{\text{intrinsic}}) - (dE_g/dP)\Delta n/3 \quad \text{--------------------------------------------(2)}$$

where $S_{\text{intrinsic}}$ is the strain caused by the indentation in intrinsic $c$-Si, $\sigma$ is the stress in the intrinsic sample, and $E_{\text{intrinsic}}$ is the apparent elastic modulus. Thus, we define the apparent elastic modulus measured in the heavily doped Si as

$$1/E_{\text{doped}} = (1/E_{\text{intrinsic}}) - (dE_g/dP)\Delta n/\sigma \quad \text{--------------------------------------------------(3)}$$

In order to determine the effect of the carrier concentration on the elastic modulus for $c$-Si, the mean compressive stress caused by nanoindentation can be estimated to be $\sigma = F/\pi R\delta$,[15] where $F$ is the load of the nanoindentation, $R$ is the radius of the indenter, and $\delta$ is the penetration depth. If taking $F = 5$ mN during loading, R ≈50 nm, and δ≈150 nm (penetration depth corresponding to 5 mN), considering the surface region which can be affected by light and ensuring purely elastic deformation is taking place alone, we estimate $\sigma \approx 20$ GPa. We measured hardness values of ~12.5 GPa for the intrinsic and lightly doped Si(100) and ~6 GPa for the heavily doped samples with the penetration depth of our nanoindentor ~150 nm (Table I). The literature reported data for the hardness are 11.9-13 GPa for undoped Si(100) (with penetration depth of 267-24 nm) and 5.1-8.7 GPa for heavily doped Si(100) (with penetration depth of 318-44 nm).[16] From Eq. (2), we can calculate the apparent elastic modulus of doped system ~174.7 GPa with is remarkably close to the estimated value of 174.6(3.9) GPa in case of $p^{++}$-Si and 178.8(2.3) GPa in $n^{++}$-Si and (as shown in Fig. 3 and in Table I) where $E_{\text{intrinsic}}$ 188.7(2) GPa. The Young modulus of bulk Si is reported be 190 by Pitersen in his seminal report predicting Si as mechanical material,[17] and is considered in the range of 169-190 taking into consideration of its anisotropic nature.[18] Thus the elastic modulus of 188.7(2) GPa measured for the undoped Si matches well with the reported values. However, for medium doped (~$1 \times 10^{17}$ cm$^{-3}$) sample the effect of carrier concentration on the apparent elastic modulus will be negligible, as estimated using Eq. (1) for electronic strain $S_e$ ~$8.5 \times 10^{-5}$ for medium doped sample (with respect to ~$8.5 \times 10^{-1}$ for heavily doped samples). Thus, the apparent elastic modulus does not change in case both $n$- and $p$-doped $c$-Si samples (Fig. 3 and in the Table I). A high doping concentration above $1 \times 10^{21}$ cm$^{-3}$ is required to observe measurable change in mechanical properties induced by charge carrier. Softening in the range of -5.3-7.5% was observed for heavily doped Si considering both $n$- and $p$-type. Role of impurity, particularly for heavily doped semiconducting materials, for influencing the elastic properties cannot be ruled out completely. However the lowering or the change in the elastic modulus due to doping alone is expected only 1-3% in the heavily doped Si as per standard "many-valleys" model,[19,20] and is also confirmed experimentally.[21-23] Thus the reduction of stiffness due to electronic origin, as studied in the present report, is pre-dominant over the impurity related contributions.

**Table I.** Elastic modulus and hardness values of *n*- and *p*-type *c*-Si

|  | Intrinsic | *n*-Si | $n^{++}$-Si | *p*-Si | $p^{++}$-Si |
|---|---|---|---|---|---|
| Elastic modulus (GPa) | 188.7 | 190.7 | 178.8 | 190.7 | 174.6 |
| Std. Dev. | 2 | 3.8 | 2.3 | 7 | 3.9 |
| Hardness (GPa) | 12.5 | 12.5 | 6 | 12.6 | 5.9 |

We must comment here that, while the stiffening is well expected in ZnO nanobelt having positive pressure coefficient of *w*-ZnO; surface effect is expected to be negligible in the study, as the highest change in the apparent elastic modulus (~ 200%) in the ZnO nanobelt by illumination occurs at a depth of 80 nm.[6] Moreover, in order to achieve 200% change in the apparent elastic modulus from the Eq. (2) using σ ≈ 5 GPa and $dE_g/dP$ ~24.5 meV/GPa for *w*-ZnO; one requires carrier injection ~ $1 \times 10^{21}$ cm$^{-3}$ in the illumination process which is unrealistic as the carrier concentration resemble the heavily doped system. Some other parameters will have to be invoked in order to explain such a high carrier injection under illumination in *w*-ZnO nanobelt. In this context, the role of thermoelasticity may be influential for the photoexcitation above the band gap where the localized heating may take place for the nanostructures with light energy getting absorbed completely. The value of thermal expansion co-efficient in the *c*-axis is reported twice than that for the *a*-axis for ZnO.[24] This may be the reason for the observed increase in the photoinduecd stiffness in ZnO nanobelt,[6] as there is a relative contraction in the basal plane while the hardness measurement is being performed along the basal plane (*ab* plane) which is normal to *c*-axis of the nanobelt.[25]

4. **Conclusion**

Thus, we can show a real effect of electronic strain in a semiconducting system where the change in mechanical property, pre-dominating over impurity effect, is truly modeled by electronic characteristics. We have also predicted the limit for the electronic effect influencing mechanical properties in semiconducting system. The study will stimulate further study of carrier concentration dependent mechanical properties in other systems and will be extremely useful for both micro electromechanical and nano-electromechanical devices where semiconductors with varying carrier concentrations are integrated for performing various mechanical functions.